\documentstyle[sprocl]{article}
\input{psfig}
\def\Journal#1#2#3#4{{#1} {\bf #2}, #3 (#4)}

\def\PR{\em Phys. Rev.}
\def\PRB{{\em Phys. Rev.} B}
\def\EPL{\em Europhys. Lett.}
\begin{document}

\title{Is the Hofstadter energy spectrum observable \\ in far-infrared
       absorption?}

\author{Vidar Gudmundsson}

\address{Science Institute, University of Iceland, Dunhaga 3,\\
         IS-107 Reykjavik, Iceland.}

\author{Rolf R.\ Gerhardts}

\address{Max-Planck-Institut f\"ur Festk\"orperforschung, 
         Heisenbergstra{\ss}e 1,\\
         D-70569 Stuttgart,  Federal Republic of Germany}

\maketitle\abstracts{The far-infrared absorption of a periodically modulated
       two-dimensional electron gas in a perpendicular constant
       magnetic field is calculated self-consistently within the 
       Hartree approximation. For vanishing modulation
       the magnetoplasmon dispersion shows simple anticrossings with
       harmonics of the cyclotron resonance, as expected. 
       For increasing modulation
       we identify intra- and intersubband magnetoplasmon modes
       charaterized by the number of flux quanta through each
       unit cell of the periodic potential.}

\noindent 
The Hofstadter energy spectrum\cite{Hofstadter76:2239,Pfannkuche92:12606} 
indirectly determines the transport prop\-er\-ties\cite{Gerhardts91:5192}
of a two-dimensional electron gas (2DEG) with a square lattice modulation. 
Recently it has been claimed that even the direct effects of the 
splitting of the Landau bands into subbands has been 
observed.\cite{Schloesser96:683} Here we discuss the possibility to directly
observe the subband splitting in optical measurements, for example, by
far-infrared (FIR) absorption, or in Raman scattering. Since screening
effects have been shown to be very important in the ground 
state\cite{Gudmundsson95:16744} we shall treat the Coulomb interaction 
between the electrons on equal footing in the excited and the ground
state by utilizing the time-dependent Hartree approximation.
Results of such calculations have been published for isolated quantum
dots\cite{Pfannkuche91:13132,Gudmundsson91:12098} and for unidirectionally
modulated 2DEG, but to our knowledge, corresponding results are not 
available for the square lattice modulation.\footnote{The collective 
excitations of a 2DEG with a two-dimensional magnetic-field modulation 
have been studied by X.\ Wu and S.\ E.\ Ulloa in Phys.\ Rev.\ B 
{\bf 47}, 10028 (1993), but without considering interaction effects 
in the ground state.}

The technical details of the ground state calculation have been published 
elsewhere.\cite{Gudmundsson95:16744} 
The square lateral superlattice with period $L$ is spanned
by the lattice vectors
${\bf R}=m{\bf l}_1+n{\bf l}_2$, where ${\bf l}_1=L\hat{\bf x}$, and
${\bf l}_2=L\hat{\bf y}$ are the primitive translations of the Bravais
lattice ${\cal B}$; $n,m\in Z$. The reciprocal lattice ${\cal R}$ is spanned by
${\bf G}=G_1{\bf g}_1+G_2{\bf g}_2$, with ${\bf g}_1=2\pi \hat{\bf y}/l_1$,
${\bf g}_2=2\pi \hat{\bf x}/l_2$, and $G_1,G_2\in Z$.
The ground-state properties
of the interacting 2DEG in a perpendicular homogeneous magnetic field
${\bf B}=B\hat{\bf z}$ and the periodic potential
$V({\bf r})=V\{\cos (g_1x)+\cos (g_2y)\}$ are calculated in the
Hartree approximation. The Hartree single-electron states $|\alpha )$
and their energies $\varepsilon_{\alpha}$ are labelled by the quantum
numbers $\{n_l,\mu ,\nu\}=\alpha$, where $n_l\geq 0$ is a Landau band index,
$\mu =(\theta_1+2\pi n_1)/p$, $\nu =(\theta_2+2\pi n_2)/q$, with
$n_1\in I_1=\{0,\cdots p-1\}$, $n_2\in I_2=\{0,\cdots q-1\}$,
$\theta_i\in [-\pi ,\pi ]$, and $pq\in N$
is the number of magnetic flux quanta through the lattice unit cell.
The Hartree potential ``felt'' by each
electron and caused by the total electronic charge density of
the 2DEG, $-en_s({\bf r})$,
and the positive neutralizing background charge, $+en_b$, together with
the external periodic potential $V({\bf r})$ do not couple states at
different points in the quasi-Brillouin zone
${\bf\theta}=(\theta_1 ,\theta_2 )\in\{ [-\pi ,\pi]\times [-\pi ,\pi]\}$.
However, states at a given point depend,
in a self-consistent way, on states in the whole zone
through $n_s({\bf r})$. The periodic potential broadens the 
Landau levels into Landau bands which due
to the comensurability conditions between $L$ and the magnetic length
$l=(c\hbar /eB)^{1/2}$, are split into $pq$
subbands. The resulting energy spectrum, the Hofstadter
butterfly\cite{Hofstadter76:2239,Pfannkuche92:12606}, retains essentially it's
complicated gap structure but is strongly reduced in symmetry by the
electron-electron interaction.\cite{Gudmundsson95:16744}
The symmetry depends on the mean electron density, or, equivalently, 
the chemical potential.

In order to calculate the FIR absorption of the 2DEG we perturb it by a
monochromatic external electric field without restricting it's dispersion
relation to that of a free propagating field. The power absorption is
calculated from the Joule heating of the self-consistent electric
field consisting of the external and the induced field.\cite{Gudmundsson96:xx}
The peaks in the absorption spectra represent absorption of the 
collective modes of the 2DEG. Single-electron excitations are strongly damped
and do not show up in the spectra.\cite{Gudmundsson96:xx}
 
In the calculation we use GaAs parameters, $m^*=0.067m_0$, and $\kappa =12.4$.
The power dissipation is made possible by retaining a small but finite 
imaginary part of the frequency in the electron 
susceptibility. 
The absorption of a homogeneous 2DEG ($V=0$) is compared in
Fig.\ 1 with the first-order dispersion in ${\bf k}$
of a magnetoplasmon,
$\omega^2=\omega_c^2+2\pi e^2n_sk/(\kappa m^*)$.
Kohn's theorem manifests itself by the fact that
only one peak is visible for ${\bf k}\rightarrow 0$.\cite{Kohn61:1242}
For finite wave vectors ${\bf k}$ the magnetoplasmon
interacts with harmonics of the
cylcotron resonance, resulting in so the called Bernstein
modes.\cite{Bernstein58:10}
To make the Hofstadter subband structure of the Landau bands discernible,
we have to resort to short periods $L=50\:$nm and strong modulation,
as is seen in Fig.\ 2. Here we use ${\bf k}L=k_1L=0.2$, and the
filling factor $\nu =1/2$. For strong modulation
and $pq=3$ the dispersion of the magnetoplasmon is relatively flat
for small values of ${\bf k}$. The structures in the absorption 
are very sensitive to the location of the chemical
potential with respect to the subbands as can be confirmed by comparing
the absorption for $\nu =1/2$ in Fig.\ 2 with the absorption 
for $\nu =5/6$ in Fig.\ 3. The inset in Fig.\ 2 shows the absorption for

\noindent
\begin{minipage}[t]{5.8cm}
      {\center{\psfig{figure=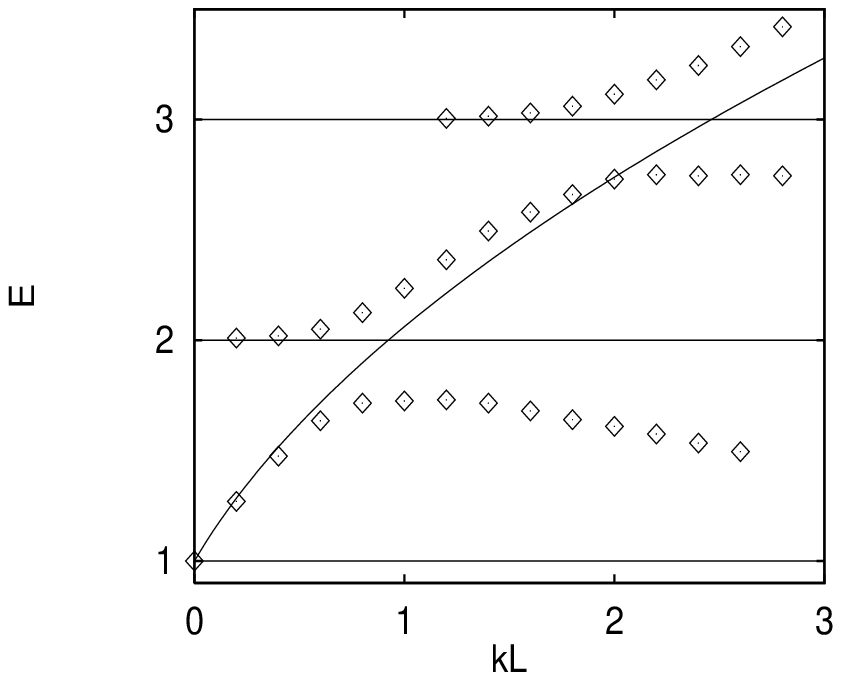,width=5.7cm}}}
\vskip 10pt
%{\footnotesize
\CaP{
Figure 1: The dispersion of the magnetoplasmon in a homogeneous 
         2DEG ($\Diamond$). 
         The energy $E$ is scaled in $\hbar\omega_c$ and the wavevector
         ${\bf k}=k_1$ is scaled with the period length $L=200\:$nm.
         $V=0$, $\hbar\omega_c=0.179\:$meV, $pq=1$, and $T=1\:$K.}
\end{minipage}
\hskip 7pt
\begin{minipage}[t]{5.8cm}
      {\center{\psfig{figure=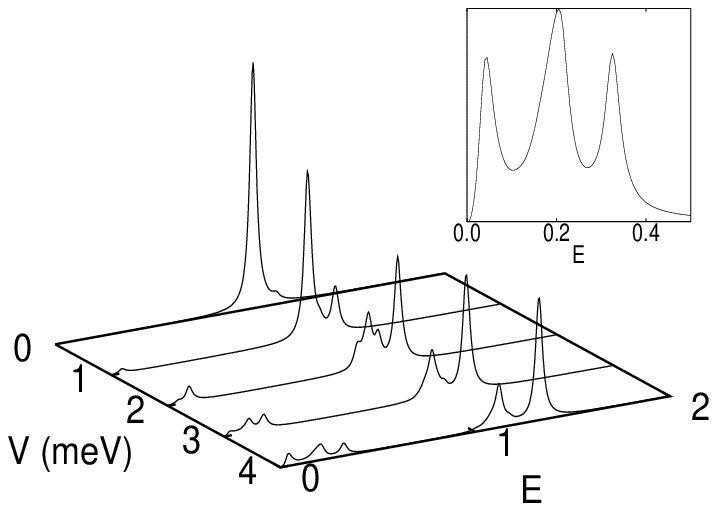,width=5.7cm}}}
\vskip 10pt
%{\footnotesize 
\CaP{
Figure 2: The absorption $P(E=\hbar\omega/\hbar\omega_c)$ 
         in arbitrary units as a function of the modulation strength $V$
         for $\nu =1/2$. 
         The inset shows the intraband absorption peaks for $V=4\:$meV.
         $k_1L=0.2$, $L=50\:$nm, $\hbar\omega_c=8.57\:$meV, $pq=3$, 
         and $T=1\:$K.}
\end{minipage}
\noindent
\begin{minipage}[t]{5.8cm}
      {\center{\psfig{figure=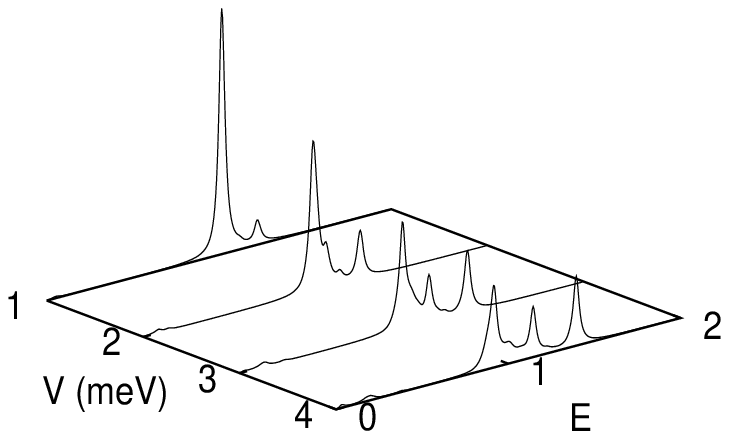,width=5.7cm}}}
\vskip 10pt
%{\footnotesize 
\CaP{
Figure 3: The absorption $P(E=\hbar\omega/\hbar\omega_c)$ 
         in arbitrary units as a function of the modulation strength $V$
         for $\nu =5/6$. $k_1L=0.2$, $L=50\:$nm, $\hbar\omega_c=8.57\:$meV, 
         $pq=3$, and $T=1\:$K.}
\end{minipage}
\hskip 7pt
\begin{minipage}[t]{5.8cm}
      {\center{\psfig{figure=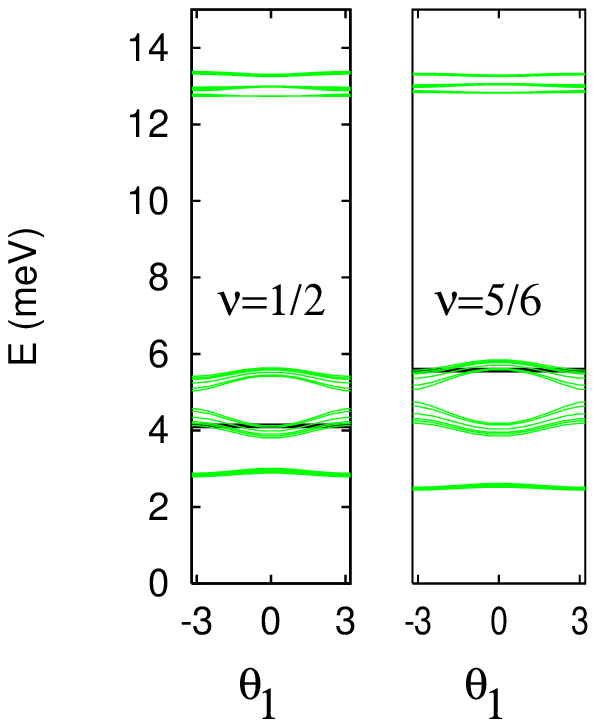,width=4.7cm}}}
\vskip 10pt
%{\footnotesize 
\CaP{
Figure 4: The Hartree dispersion of the two
         lowest Landau bands in the quasi-Brillouin zone (each band
         is split into three subbands). The chemical potential is
         indicated by a solid horizontal line.
         $L=50\:$nm, $V=4\:$meV, $\hbar\omega_c=8.57\:$meV, 
         $pq=3$, and $T=1\:$K.}
\end{minipage}

\vskip 10pt
\noindent
{\it three} intra-Landau band magnetoplasmon modes.
For $\nu =1/2$ in Fig.\ 2 only two subbands are occupied, 
or partially
occupied, as is illustrated in Fig.\ 4. 
We have thus {\it two} absorption peaks for inter-Landau band magnetoplasmons.
Fig.\ 4 also shows that in the case of $\nu =5/6$ we can have {\it three}
absorption peaks for interband magnetoplasmons. The strong screening
effects of partially filled subbands weakens any selections rules that
would apply in the case of a noninteracting 2DEG. This is especially
evident by comparing the results here to the case of $\nu =1$, where 
only one absorption peak is present
in the spectrum due to the very weak screening of filled Landau bands. 

In summary, the FIR-absorption of a 2DEG in a short-period strongly modulated
square lateral superlattice reflects the underlying Hofstadter
subband structure. The structure can be seen in the absorption
due to either intra- or inter-Landau band magnetoplasmons by 
tuning the density or the
filling factor within the lowest Landau band.
Due to the depolarization shift of the peaks an exact information about
the Hofstadter butterfly can not be extracted from the absorption spectra.
The intra-Landau band magnetoplasmon peaks in the 
absorption have weak oscillator
strengths compared to the interband absorption and are situated in a region
on the energy scale that is difficult to detect with present day
FIR technology.

\section*{Acknowledgments}
This research was supported in part by the Icelandic
Natural Science Foundation, the University of Iceland Research Fund,
and a NATO collaborative research Grant No. CRG 921204.

\end{document}